\documentclass[apj]{emulateapj}
\usepackage{natbib}
\usepackage{adjustbox}

\shorttitle{Expansion of the hydrogen-poor Knots in A30 and A78}
\shortauthors{X.\ Fang et al.}

\begin{document}

\title{
Expansion of hydrogen-poor knots in the Born-Again Planetary Nebulae 
A30 and A78\footnotemark[$\ast$]}

\footnotetext[$\ast$]{Based on observations made with the NASA/ESA
\emph{Hubble Space Telescope}, obtained at the Space Telescope Science
Institute, which is operated by the Association of Universities for Research
in Astronomy, Inc., under NASA contract NAS\,5-26555.  These observations 
are associated with program \#12935.
}

\author{
X.\ Fang$^{1}$, 
M.~A.\ Guerrero$^{1}$, 
R.~A.\ Marquez-Lugo$^{1}$, 
J.~A.\ Toal\'a$^{1}$, 
S.~J.\ Arthur$^{2}$,
Y.-H.\ Chu$^{3}$\footnotemark[$\dagger$], 
W.~P.\ Blair$^{4}$,\\
R.~A.\ Gruendl$^{3}$,
W.-R.\ Hamann$^{5}$, 
L.~M.\ Oskinova$^{5}$, 
and 
H.\ Todt$^{5}$ \\
}
\affil{
$^{1}$Instituto de Astrof\'{i}sica de Andaluc\'{i}a (IAA-CSIC), Glorieta 
de la Astronom\'{i}a s/n, E-18008 Granada, Spain\\
$^{2}$Centro de Radioastronom\'{i}a y Astrof\'{i}sica, Universidad Nacional
Aut\'{o}noma de M\'{e}xico, Campus Morelia, 58090 Morelia, Mexico\\
$^{3}$Department of Astronomy, University of Illinois, 1002, West Green 
Street, Urbana IL 61801, USA\\
$^{4}$Department of Physics and Astronomy, Johns Hopkins University, 
Baltimore, MD 21218, USA\\
$^{5}$Institute for Physics and Astronomy, Universit\"{a}t Potsdam, 14476 
Potsdam, Germany\\
}
\email{fangx@iaa.es}

\footnotetext[$\dagger$]{Now at the Institute of Astronomy and Astrophysics, 
Academia Sinica (ASIAA), Taipei 10617, Taiwan.}

\begin{abstract}
We analyze the expansion of hydrogen-poor knots and filaments in the 
born-again planetary nebulae A30 and A78 based on \emph{Hubble Space 
Telescope} (\emph{HST}) images obtained almost 20 yr apart.  The proper 
motion of these features generally increases with distance to the central 
star, but the fractional expansion decreases, i.e., the expansion is not 
homologous.  As a result, there is not a unique expansion age, which is 
estimated to be 610--950~yr for A30 and 600--1140~yr for A78.  The knots and 
filaments have experienced complex dynamical processes: the current fast 
stellar wind is mass loaded by the material ablated from the inner knots; 
the ablated material is then swept up until it shocks the inner edges of the 
outer, hydrogen-rich nebula.  The angular expansion of the outer filaments 
shows a clear dependence on position angle, indicating that the interaction 
of the stellar wind with the innermost knots channels the wind along 
preferred directions.  The apparent angular expansion of the innermost knots 
seems to be dominated by the rocket effect of evaporating gas and by the 
propagation of the ionization front inside them.  
Radiation-hydrodynamical simulations show that a single ejection of material 
followed by a rapid onset of the stellar wind and ionizing flux can reproduce 
the variety of clumps and filaments at different distances from the central 
star found in A30 and A78.
\end{abstract}


\section{Introduction} \label{section1}

Born-again planetary nebulae (PNe) are believed to have experienced 
a \textit{very late thermal pulse} \citep[VLTP;][]{Iben1983} while  
the star was descending the white dwarf cooling track. 
During this event, the remnant stellar helium envelope reaches 
the critical mass to ignite its fusion into carbon and oxygen 
\citep[e.g.,][]{herwig05,miller06,lm06}; 
the sudden increase of pressure leads to the ejection of the newly 
processed material and, as the stellar envelope expands, its 
temperature decreases and the star returns in the Hertzsprung-Russell 
(HR) diagram to the locus of the asymptotic giant branch (AGB) stars.  
Soon after, the contraction of the envelope will increase the stellar 
effective temperature, boosting the UV flux and giving rise to a new 
fast stellar wind.  

So far, the only \textit{bona-fide} born-again PNe are Abell\,30 (aka 
A30), Abell\,58 (aka A58, Nova\,Aql\,1919,), Abell\,78 (aka A78), 
and V4334 Sgr (aka Sakurai's object).  
Among them, A30 and A78 are the more evolved ones, with large 
limb-brightened, H-rich outer nebulae 
surrounding the H-poor, irregular-shaped structures that harbour the 
``cometary'' knots in the innermost regions 
\citep{Jacoby1979,Meaburn1996,Meaburn1998}. 
Detailed \emph{Hubble Space Telescope} (\emph{HST}) images in the 
[O~{\sc iii}] emission line of the central regions have revealed 
equatorial rings and compact polar outflows in the central regions 
of both PNe \citep{Borkowski1993,bor95}. 
The dynamics are revealing: while the outer nebulae show shell-like 
expansions with velocities of 30--40~km~s$^{-1}$, the H-poor clumps 
present complex structures, with velocity spikes $\gtrsim$200 km~s$^{-1}$
\citep{Meaburn1996,chu97,Meaburn1998}.

The morphology and kinematics of the H-poor knots unveil rich dynamical 
processes in the nebulae.  The material photoevaporated from the knots 
by the stellar radiation is swept up downstream by the fast stellar wind, 
which is otherwise mass loaded and slowed down \citep{pittard05}.  The 
interactions are complex, resulting in sophisticated velocity structures 
\citep[e.g.,][]{Steffen2004}, as well as X-ray-emitting hot gas 
\citep{chu95,Guerrero2012,Toala2014}.


To gain further insights into the dynamics of the H-poor knots and filaments 
in A30 and A78, we study their proper motions using \emph{HST} images 
taken $\sim$20~yr apart.  
These observations and the data analysis are described in 
Section~\ref{section2}.  
The results are presented in Section~\ref{section3}.  
In Section~\ref{section4} we discuss our findings and present basic 
radiation-hydrodynamic simulations. 
Finally, we present our conclusions in Section~\ref{section5}.

\begin{figure*}[!t]
\begin{center}
\includegraphics[width=2.08\columnwidth,angle=0]{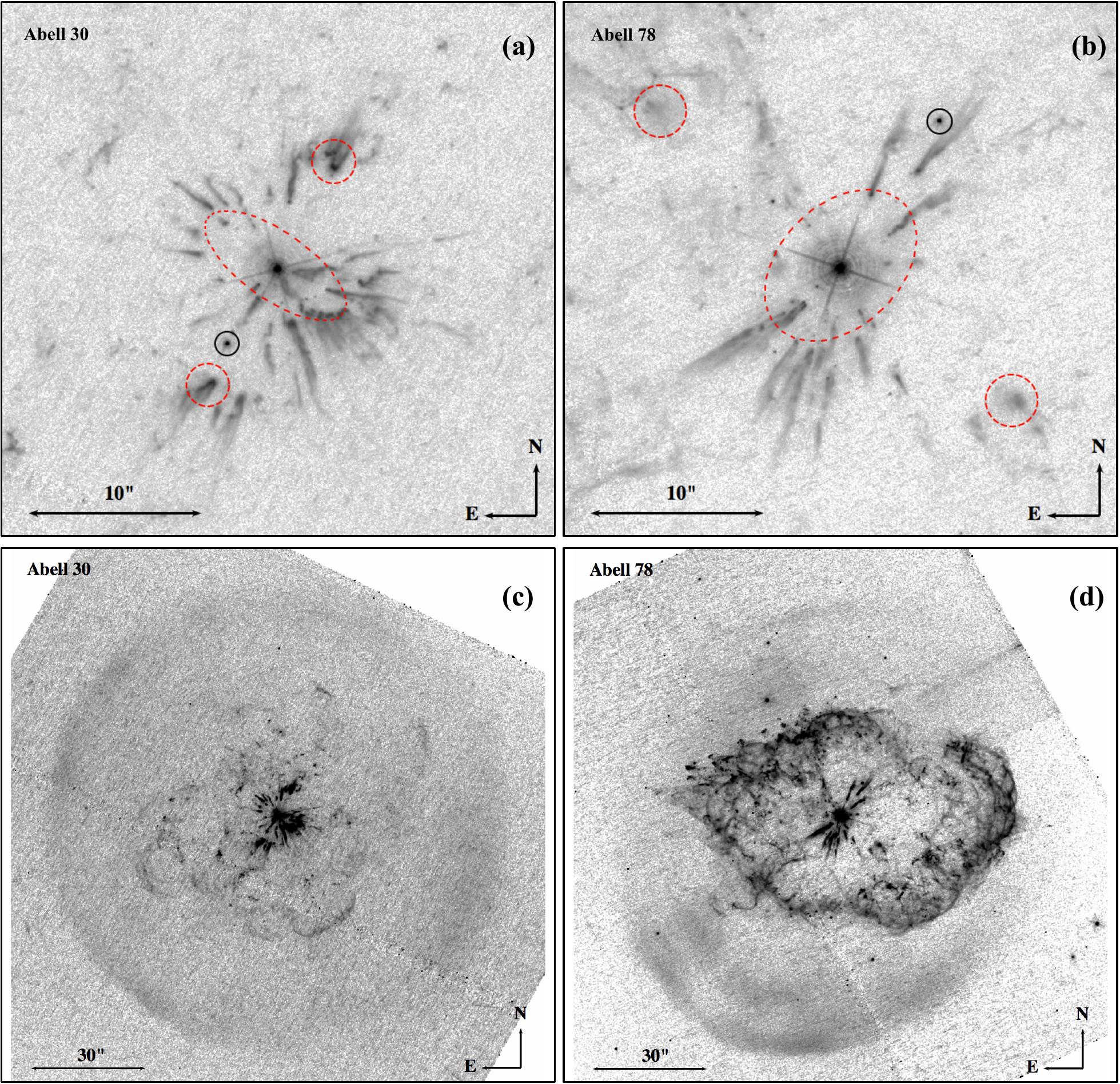}
\caption{
Negative grey-scale \emph{HST} WFC3 [O~{\sc iii}] narrowband images 
of A30 and A78: 
The inner regions of A30 and A78 are shown in panels \emph{a} and \emph{b}, 
respectively, where the equatorial disks and bipolar knots described by 
\cite{Borkowski1993,bor95} are delineated with red dashed ellipses and 
circles.  
Prominent field stars in the inner regions are marked with black circles. 
Panels \emph{c} and \emph{d} show the H-poor outer filaments, and the 
smooth, H-rich outer nebulae of A30 and A78, respectively.  The H-poor 
outer filaments have a petal-like appearance in A30, and a spindle shape 
in A78.  
}
\label{fig1}
\end{center}
\end{figure*}

\begin{figure*}[!t]
\begin{center}
\includegraphics[width=2.3\columnwidth,angle=-90]{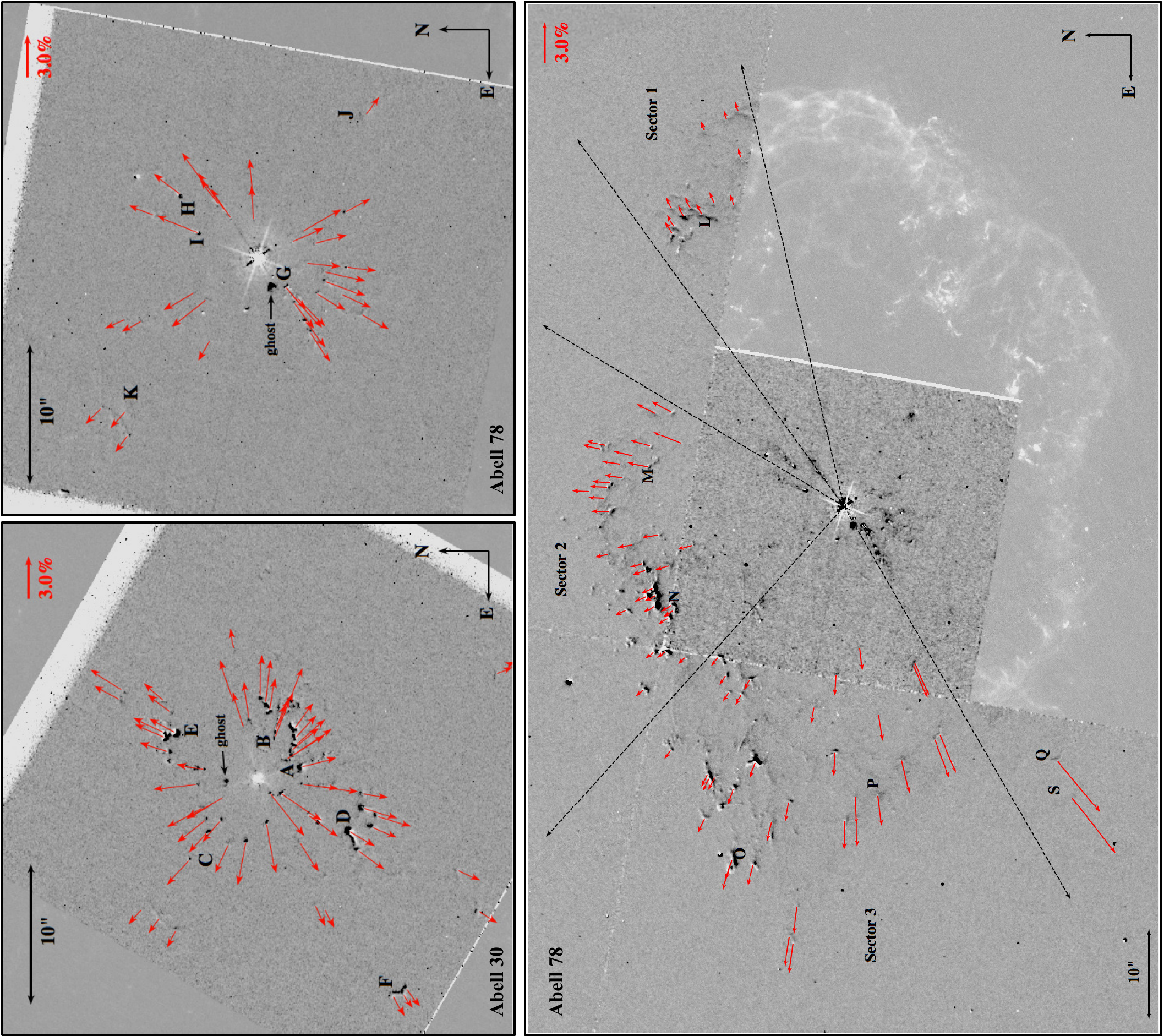}
\caption{
Residual images showing the expansion of the H-poor knots and filaments 
in the inner regions of A30 (\emph{top-left}) and A78 (\emph{top-right}), and 
in the outer region of A78 (\emph{bottom}).  
The 2012/13 WFC3 data are shown as white while the 1994/95 WFPC2 data 
are shown as black.  
The ghosts originating from the \emph{HST} WFPC2 images are marked.  
Expansion is evident for the innermost knots of both PNe and for the outer 
filaments of A78.  
The red arrows overlaid show the directions of the expansion, with lengths 
proportional to the values of fractional expansion $\delta_\mathrm{exp}$ as 
defined by Eq.~\ref{eq1}.  
A 3.0\% fractional expansion arrow is indicated on the upper-right corner 
of each panel.  
Knots and filaments listed in Table~\ref{expansion} are labeled.  
The black dashed arrows, dividing the outer region of A78 into the 
three sectors defined according to Figure~\ref{fig4}, are described 
in Section~\ref{section3}.
}
\label{fig2}
\end{center}
\end{figure*}

\section{Observations and Data Analysis} \label{section2}

New \emph{HST} WFC3 F502N narrowband images of A30 and A78 (PI: 
M.A.\ Guerrero, Prop.\ ID 12935) were obtained on 2013 March 21 and 2012 
November 22, respectively.  
The images were reduced following standard pipeline procedures.  
The first-epoch \emph{HST} WFPC2 F502N narrowband images of A30 and 
A78 (PI: J.P.\ Harrington, Prop.\ ID 5404 and 5864) were downloaded 
from MAST\footnote{URL, http://archive.stsci.edu}, the Mikulski 
Archive for Space Telescope at the Space Telescope Science Institute.  
These images were obtained on 1994 March 6 and 1995 July 11, 
respectively.  Thus the time-lapse between the two epochs is 19.04~yr 
for A30 and 17.36~yr for A78. 
The properties of the \emph{HST} WFC3 and WFPC2 images of A30 and A78 
are summarized in Table~\ref{tab:compare}. 
Basically, both the new WFC3 and archival WFPC2 images map exclusively 
the [O~{\sc iii}] emission line.  
The main differences arise from the pixel scale, which is similar for 
WFC3-UVIS and WFPC2-PC, but $\approx$2.5 times larger for WFPC2-WF than 
for WFC3-UVIS.

\begin{figure*}
\begin{center}
\includegraphics[width=2.05\columnwidth,angle=0]{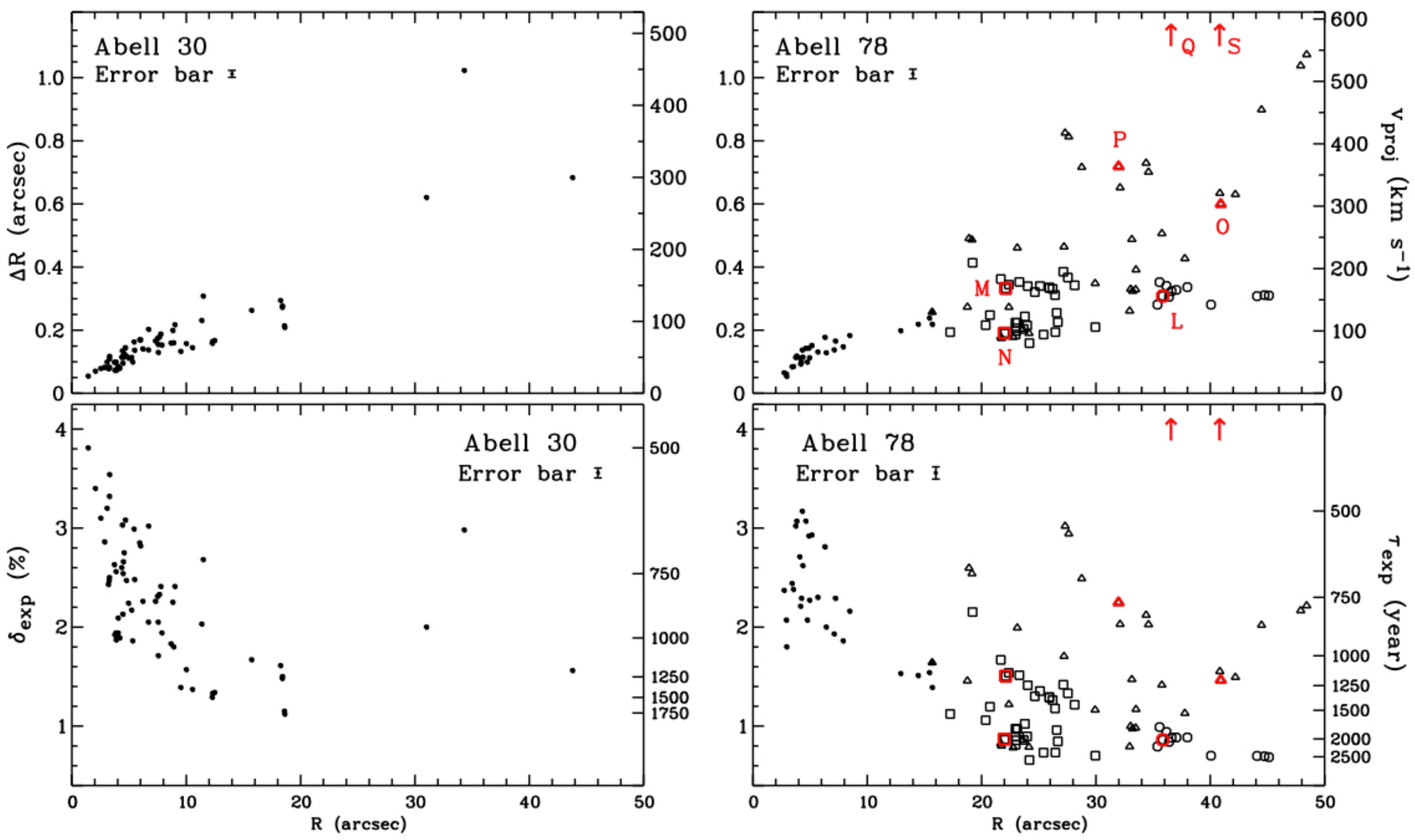}
\caption{
Radial variations of the angular expansion ${\Delta}R$ and sky-projected 
velocity $v_\mathrm{proj}$ (\emph{top}), and the fractional expansion 
$\delta_{\rm exp}$ and apparent expansion age $\tau_{\rm exp}$ 
(\emph{bottom}) of the H-poor knots and filaments in A30 (\emph{left}) and 
A78 (\emph{right}).  
For each PN, the typical uncertainty in ${\Delta}R$ and $\delta_{\rm exp}$, 
$\lesssim$1\%, is indicated with an error bar.  
Circles, squares and triangles in the \emph{right} panels correspond 
to the outer knots and filaments of A78 in Sectors~1, 2 and 3 
(see Figure~\ref{fig2}-\emph{bottom} and Figure~\ref{fig4}), respectively.  
The outer features labeled in Figure~\ref{fig2}-\emph{bottom} and listed 
in Table~\ref{expansion} are highlighted in red.  
The positions of the features ``Q'' and ``S'', with 
$v_\mathrm{proj}\simeq$650\,km~s$^{-1}$ and 770\,km~s$^{-1}$, and 
$\tau_\mathrm{exp}\simeq$360\,yr and 330\,yr, respectively, lay outside 
the panels, as indicated by red arrows. 
}
\label{fig3}
\end{center}
\end{figure*}


\begin{table*}
\begin{center}
\caption{Details of the \emph{HST} WFPC2 and WFC3 observations}
\label{tab:compare}
\begin{tabular}{lcclccc}
\hline
\hline
\multicolumn{1}{l}{Instrument} &
\multicolumn{1}{c}{Camera}     &
\multicolumn{1}{c}{Pixel Size} &
\multicolumn{1}{c}{Date~~~~} &
\multicolumn{3}{c}{\underline{~~~~~~~~~~~~~~Filter~~~~~~~~~~~~~~}} \\
\multicolumn{1}{c}{} &
\multicolumn{1}{c}{} &
\multicolumn{1}{c}{} &
\multicolumn{1}{c}{} &
\multicolumn{1}{c}{Name} &
\multicolumn{1}{c}{$\lambda_{\rm c}$} &
\multicolumn{1}{c}{Bandwidth} \\
\multicolumn{1}{c}{} & 
\multicolumn{1}{c}{} & 
\multicolumn{1}{c}{} &
\multicolumn{1}{c}{} &
\multicolumn{1}{c}{} &
\multicolumn{1}{c}{(\AA)} &
\multicolumn{1}{c}{(\AA)} \\
\hline
\multicolumn{7}{c}{~~A30} \\
\hline
WFPC2 & PC/WF &  0\farcs0454/0\farcs0996~~~~ & March 6, 1994~~~~     & F502N & 5013 & 27 \\
WFC3   & UVIS    &  0\farcs0396~~~~                     & March 21, 2013~~~~   & F502N & 5010 & 65 \\
\hline
\multicolumn{7}{c}{~~A78} \\
\hline
WFPC2 & PC/WF &  0\farcs0454/0\farcs0996~~~~ & July 11, 1995~~~~           & F502N & 5013 & 27 \\
WFC3   & UVIS    &  0\farcs0396~~~~                     & November 22, 2012~~~~ & F502N & 5010 & 65 \\
\hline
\end{tabular}
\end{center}
\end{table*}

The new \emph{HST} WFC3 images of A30 and A78 are presented in 
Figure~\ref{fig1}, where their inner regions are shown in the top 
panels and the complete view of the nebulae in the bottom panels.  
These new images confirm the morphological 
features detected in previous \emph{HST} imagery \citep{Borkowski1993,bor95}: 
a series of compact knots with cometary tails pointing away from the central 
stars (CSPNe) are surrounded by the outer filamentary irregular (petal-like) 
structures.  The inner H-poor knots are mostly located on the equatorial 
disks (red dashed ellipses in the top panels of Figure~\ref{fig1}) and bipolar
regions (red dashed circles in Figure~\ref{fig1}).  
The petal-like structures of A30 and intermediate spindle-shaped shell 
of A78 are completely mapped by these images.  
The round and oval-shaped outer shells of A30 and A78, respectively, 
are also registered by these \emph{HST} WFC3 images.

In order to investigate the proper motions of the knots and filaments in A30 
and A78, we have compared the images obtained at the two epochs. 
The WFC3 and WFPC2 images were first carefully aligned along the sky 
coordinates with the pixel scale of the WFPC2 cameras rebinned to that of the
WFC3 camera using standard {\sc iraf}\footnote{{\sc iraf} is distributed by 
the National Optical Astronomy Observatory, which is operated by the 
Association of Universities for Research in Astronomy, Inc., under 
cooperative agreement with the National Science Foundation.} routines. 
Then the background emission was subtracted by adopting a count
level computed from regions outside the outer nebulae.  The images were then 
scaled to the same level using the count peaks of the brightest and more 
compact knots, and subsequently subtracted.  The residual images in 
Figure~\ref{fig2} clearly show the expansion of the innermost H-poor knots 
and irregular outer filaments.  
A typical angular expansion of $\gtrsim$0\farcs1 was found for the knots in 
the innermost regions, and up to 1\arcsec\ for the features in the outermost 
regions. 
As in \citet{reed99}, \citet{Guerrero2012}, or \citet{odell13}, the older 
WFPC2 images were magnified and subtracted from the most recent WFC3 
images.  By inspecting the residual images, we found that there is not a unique 
magnifying factor for the different features.

%
%

It is pertinent to estimate whether geometric distortion affects our 
analysis of the \emph{HST} WFC3 and WFPC2 images.  
The WFC3 UVIS detector is not strictly aligned with the optical axis of the 
telescope, which contributes a linear distortion.  
Moreover, the plate scale varies across the detector, adding a non-linear 
distortion component which is maximal at the corners of the detector.  
Both components are corrected by the \emph{HST} standard WFC3 pipeline, 
with a typical accuracy of 0.1 pixels, i.e., $\sim$4~mas, at the detector 
edges and much smaller in the central regions according to the WFC3 Data 
Handbook\footnote{URL, 
http://www.stsci.edu/hst/wfc3/documents/handbooks}.  
As for the WFPC2 images, the current geometric distortion correction 
procedures remove most of the linear distortion and essentially all 
non-linear distortion according to the WFPC2 Data Handbook\footnote{URL, 
http://www.stsci.edu/hst/wfpc2/documents/handbook}.  
Therefore, the largest uncertainties in the geometric correction at the 
detector edges, $\sim$4~mas, are still smaller than the typical angular 
expansions ($\gtrsim$100~mas) detected in the central regions of the 
residual images of A30 and A78.  
Inspection of field stars in the residual images further confirms 
there are no significant systematic deviations in their positions.  
We are thus confident that geometric distortions of the images 
do not substantially affect the results presented in this paper. 
The only noticeable effect is a well known ghost that affects the WFPC2 
images which is marked in Figure~\ref{fig2}.

To quantify the expansion of the different structural components of A30 and 
A78, we measured the radial angular distances of the knots and filaments in 
the WFPC2 ($R_\mathrm{WFPC2}$) and WFC3 ($R_\mathrm{WFC3}$) images.  
About 60 features were analyzed in A30, including $\sim$50 in the central region 
($R\leq$15\arcsec), and 104 knots and filaments in A78, including 30 in the 
central region ($R\leq$13\arcsec).  The petal-like structure of A30 is in 
nature fainter than that of A78 and is only partially covered by the WFPC2 
camera, thus its expansion can not be accurately assessed.

The angular expansion ($\Delta R=R_\mathrm{WFC3} - R_\mathrm{WFPC2}$) 
and the velocity component on the plane of the sky of the different features, 
computed as 
\begin{equation}
\label{eq3}
v_\mathrm{proj} = d \times \frac{\Delta R}{\Delta t}, 
\end{equation}
are plotted against the radial distance in Figure~\ref{fig3}-{\it top}, 
assuming a distance $d$=1.8~kpc for A30 \citep[as adopted by][]{Guerrero2012}
and 1.4~kpc for A78 \citep[][]{Toala2014}, and a time-lapse 
between the two observations $\Delta{t}$=19.04~yr for A30 and 17.36~yr for 
A78.  These plots clearly illustrate a general trend that the angular 
expansion increases with radial distance.  
There is not, however, a tight linear correlation, as is apparent from the 
scatter seen for the outer knots and filaments.  
The fractional expansion is computed as  
\begin{equation}
\label{eq1}
\delta_{\rm exp} = \frac{R_{\rm WFC3} - R_{\rm WFPC2}}{R_{\rm WFPC2}}, 
\end{equation}
which is equivalent to the ratio between the current and the average 
expansion velocities and is independent of the inclination of motion with 
respect to the line of sight.  The values of $\delta_\mathrm{exp}$ of the 
knots and filaments are shown in Figure~\ref{fig2} by arrows.  Expansion of 
the southwest outer region of A78 was not studied because this quadrant was 
not covered by the WFPC2 camera.  The apparent expansion ages were estimated 
by 
\begin{equation}
\label{eq2}
\tau_{\rm exp}~({\rm yr}) \approx \Delta{t}/\delta_{\rm exp}. 
\end{equation}
The distributions of $\delta_\mathrm{exp}$ and $\tau_\mathrm{exp}$ with 
radial distance are shown in the lower panels of Figure~\ref{fig3}.

\section{Results} \label{section3}

\subsection{General Expansion} \label{section3:a}

The distributions of $\Delta\,R$, $v_\mathrm{proj}$, $\delta_\mathrm{exp}$, 
and $\tau_\mathrm{exp}$ shown in Figure~\ref{fig3} confirm a 
non-homologous expansion for the knots and filaments of A30 and A78.  
The innermost knots show an increase in angular expansion with radial 
distance (Figure~\ref{fig3}-\emph{top}) that tends to flatten out as the radial 
distance increases.  Correspondingly, the fractional expansion decreases 
and the expansion age increases with radial distance (Figure~\ref{fig3}-\emph{bottom}).  
The outer knots and filaments show a large dispersion, but 
in general confirm the flattening of the angular expansion with radial distance.

The proper motions ($\mu = \Delta R/\Delta t$), $\delta_\mathrm{exp}$, and 
$\tau_\mathrm{exp}$ of some prominent features marked in Figure~\ref{fig2} 
are listed in Table~\ref{expansion}.  
Overall, the features in the equatorial rings (ERs) have larger 
$\delta_\mathrm{exp}$ than the polar outflows (POs), and both 
ER and PO features have larger $\delta_\mathrm{exp}$ than the 
few filamentary features within 20\arcsec\ of the CSPNe (e.g., 
feature ``F'' in Table~\ref{expansion}).  
The statistics of all features implies an averaged $\delta_\mathrm{exp}$ 
of 2.6$\pm$0.5\% for the ER of A30, 2.1$\pm$0.1\% for its PO, and only 
1.6$\pm$0.5\% for the filamentary features.  
Similarly, the ER and PO of A78 have averaged $\delta_\mathrm{exp}$ of 
2.5$\pm$0.4\% and 1.5\%, respectively.

\begin{table}
\begin{center}
\caption{Expansion of prominent H-poor knots in A30 and A78}
\label{expansion}
\begin{tabular}{lcrrrcr}
\hline\hline
\rule[-1mm]{0mm}{5.0mm}
ID\tablenotemark{a} & 
Feature\tablenotemark{b} & 
$R_{\rm WFPC2}$ & 
$R_{\rm WFC3}$ & 
$\mu$~~~~~~ & 
$\delta_{\rm exp}$ & 
$\tau_{\rm exp}$ \\
  &
  &
  &  & (mas~yr$^{-1}$) & (\%) & (yr) \\
\hline
\multicolumn{7}{c}{A30} \\
\hline
A & ER &  3\farcs206 &  3\farcs286 &  4.20~~~~ & 2.50 &  780 \\
B & ER &  3\farcs178 &  3\farcs284 &  5.57~~~~ & 3.32 &  590 \\
C & ER &  5\farcs218 &  5\farcs315 &  5.09~~~~ & 1.86 & 1020 \\
D & PO &  7\farcs385 &  7\farcs536 &  7.93~~~~ & 2.05 &  940 \\
E & PO &  6\farcs569 &  6\farcs702 &  6.99~~~~ & 2.02 &  950 \\
F & OF & 17\farcs955 & 18\farcs245 & 15.23~~~~ & 1.61 & 1200 \\
\hline
\multicolumn{7}{c}{A78} \\
\hline
G & ER &  2\farcs872 &  2\farcs931 &  3.40~~~~ & 2.07 & 850 \\
H & ER &  7\farcs058 &  7\farcs220 &  9.33~~~~ & 2.29 & 780 \\
I & ER &  4\farcs498 &  4\farcs636 &  7.95~~~~ & 3.07 & 600 \\
J & PO & 12\farcs734 & 12\farcs929 &  11.23~~~~ & 1.53 & 1150 \\
K & PO & 14\farcs244 & 14\farcs459 &  12.38~~~~ & 1.51 & 1160 \\
L & OF & 36\farcs315 & 36\farcs608 &  16.88~~~~ & 0.88 & 1960 \\
M & OF & 21\farcs820 & 22\farcs103 &  16.30~~~~ & 1.50 & 1150 \\
N & OF & 21\farcs797 & 21\farcs986 &  10.89~~~~ & 0.87 & 1990 \\
O & OF & 40\farcs323 & 40\farcs914 &  34.04~~~~ & 1.46 & 1180 \\
P & OF & 31\farcs277 & 31\farcs980 &  40.49~~~~ & 2.25 &  770 \\
Q & OF & 34\farcs867 & 36\farcs568 &  97.98~~~~ & 4.88 &  360 \\
S & OF & 38\farcs799 & 40\farcs814 & 116.10~~~~ & 5.19 &  330 \\
\tableline
\end{tabular}
\tablenotetext{{\rm a}}{
The knot ID is labeled in Figure~\ref{fig2}.  }
\tablenotetext{{\rm b}}{
ER=equatorial ring, PO=polar outflow, OF=outer filament.  
}
\end{center}
\end{table}

\begin{figure}[!t]
\begin{center}
\includegraphics[width=0.65\columnwidth,angle=-90]{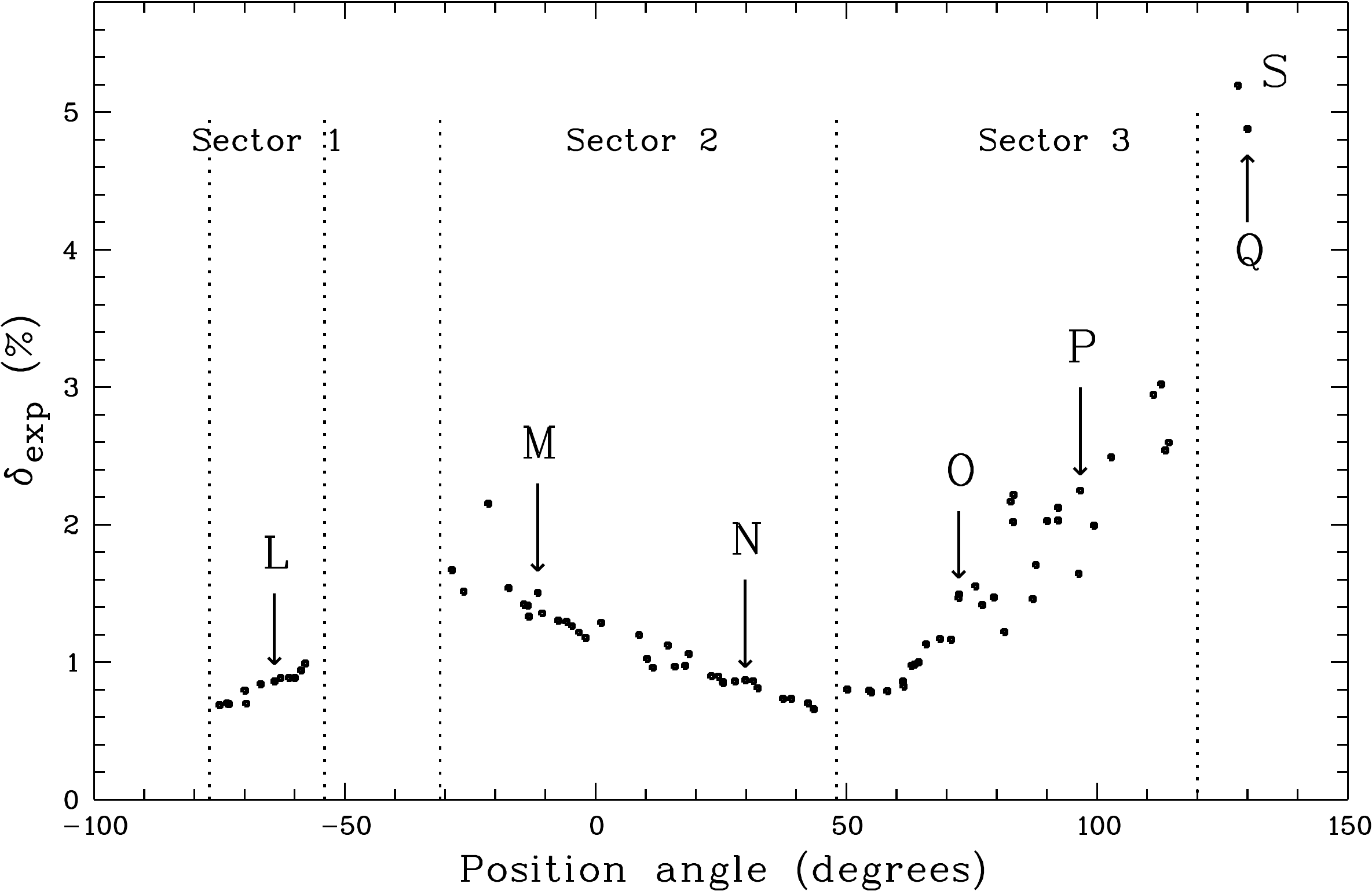}
\caption{
Variation with position angle of the fractional expansion $\delta_{\rm exp}$ 
of the outer knots and filaments of A78.  Features are separated into 
three angular sectors, depending on the mean value and trend of 
$\delta_{\rm exp}$.  The outer features labeled in 
Figure~\ref{fig2}-\emph{bottom} and Figure~\ref{fig3}-\emph{right} are 
marked, and their expansion information is given in Table~\ref{expansion}. 
The typical uncertainty in $\delta_{\rm exp}$ is $\lesssim$1\%. 
}  
\label{fig4}
\end{center}
\end{figure}

As for the outermost filaments, there is only useful information for 
those of A78.  Inspection of the variation of $\delta_\mathrm{exp}$ 
with position angle (PA) reveals a notable correlation as shown in 
Figure~\ref{fig4}.  We have used this plot to define three sectors in 
the outer region of A78, as shown in Figure~\ref{fig2}-\emph{bottom}.  
Actually, these sectors can be associated with three different morphological 
features of the spindle-shaped inner shell of A78: 
sector~1 with the westernmost spindle-bottom,  
sector~2 with a northern blister, and 
sector~3 with the easternmost arrow-shaped spindle-top.  
The spread of $\delta_\mathrm{exp}$ with radial distance in 
Figure~\ref{fig3}-\emph{right} is then clarified, as it is shown in 
Figure~\ref{fig4} that the features in each sector follow different 
trends. 
We note that there is a gap between sectors 1 and 2 (from 
PA~$\approx$\,$-$54$\degr$ to $-$30$\degr$) in Figure~\ref{fig4}, as 
the filaments and knots in that region are very diffuse and extremely 
faint.  After image subtraction (WFC3 $-$ WFPC2), we could not find any 
obvious features in that region of the residual image (see 
Figure~\ref{fig2}-\emph{bottom}).  Figure~\ref{fig1}-\emph{bottom-right} 
has been scaled to enhance these features.  
We found that there is no obvious correlation between $\delta_\mathrm{exp}$ 
and the PA for the inner knots of A30 and A78, as will be discussed in 
Section~\ref{section3:b}

The large variation of $\delta_\mathrm{exp}$ with position angle reveal
the anisotropic expansion of the spindle-shaped shell in A78.  
It is interesting to note some features in sector~3 of A78 with large
$\delta_\mathrm{exp}$ (the triangle symbols in 
Figure~\ref{fig3}-\emph{bottom-right}). 
Some of them have the appearance of compact knots (e.g., feature ``O'' 
in Figure~\ref{fig2}-\emph{bottom}) and bow-shock structures, suggesting 
that they are probably coasting away ballistically and have not been slowed 
down by the outer nebular shell yet.  
Very remarkably, the filamentary features ``Q'' and ``S'' have 
fractional expansions $\approx$5\% (Table~\ref{expansion}) that 
imply projected velocities on the plane of the sky $>$600~km~s$^{-1}$.  
These are among the highest expansion velocities associated with 
collimated outflows in PNe \citep{guerrero2002}, 
such as the knotty outflow with velocities over 500~km\,s$^{-1}$ of the 
young PN MyCn\,18 \citep{bryce1997}.

The polar outflows of A30 and A78 show intermediate values of 
$\delta_\mathrm{exp}$. 
In general, they follow the trend from high $\delta_\mathrm{exp}$ for the
equatorial knots to low $\delta_\mathrm{exp}$ for the outer filaments.
The polar outflows of A30, closer to the equatorial knots, have
$\delta_\mathrm{exp}$ values similar to the equatorial values, whereas 
the polar outflows of A78, farther from the equatorial knots, are more
like the outer filamentary features along the first half of sector~2. 
This may imply that the bipolar outflows slow down as they travel away,
similar to those found in NGC\,6543 \citep{reed99}.



\subsection{3D Kinematics of the Inner H-poor Knots} 
\label{section3:b}


The mean velocity projected on the plane of the sky of the polar
knots measured from the two epochs of observations is 62~km\,s$^{-1}$ 
for A30 and 78~km\,s$^{-1}$ for A78 (see Eq.~1).  
Meanwhile, the spatio-kinematical study of \citet{Meaburn1996} reveals a 
systemic radial velocity for the A30 Northern polar knot 
\citep[i.e., the knot J3;][]{Jacoby1979} of $-$28~km\,s$^{-1}$, 
which is generally consistent with the measurements of \citet{chu97}. 
Similarly, \citet{Meaburn1998} derived a systemic radial velocity of 
98~km\,s$^{-1}$ for the Southern polar knot of A78.  
These results are summarized in Table~\ref{tab:summary}, together with 
the velocity modulus (${\vert}v\vert$) and inclination angle ($i$) of 
the velocity vector of the polar outflows with respect to the line of 
sight.
%
%

As for the equatorial rings, the averaged values of $\delta_\mathrm{exp}$ 
imply expansion velocities of 56~km\,s$^{-1}$ for A30 and 30~km\,s$^{-1}$ 
for A78 (Table~\ref{tab:summary}).
%
Assuming the equatorial rings to be circular and the polar outflows 
perpendicular to them, the newly derived inclination angles imply 
ellipses with the minor-to-major axis ratios of 0.4 for A30 and 0.7 
for A78.  
While the axis ratio seems to fit the overall spatial distribution 
of the knots at the equatorial ring of A78, it seems inconsistent 
for A30.  
Either the polar outflows of A30 do not move perpendicular to its 
equatorial ring, or the latter does not expand isotropically in a 
plane.

Since a circular, two-dimensional (2D) and homogeneously-expanding 
ring would produce a characteristic signature in the azimuthal 
variation of $\mu$, whereas $\delta_\mathrm{exp}$ would remain constant, 
we have analyzed these quantities for $30$ knots in the equatorial ring 
of A30 and $20$ in that of A78.  
It turns out that the variations of $\mu$ and $\delta_\mathrm{exp}$ 
do not seem to depend on the azimuthal angle.
Indeed, the knots along the same direction from the CSPN show 
noticeable discrepancy in the values of $\mu$, which is in contrast with 
what we originally expected, i.e., $\mu$ should be clearly correlated 
with radial distance from the CSPN.  
We conclude that the apparent expansion of individual knots does not 
fulfill the expectations for an equatorial ring expanding as a whole.  
The geometry (non-planar distribution of knots) or dynamics (non-circular 
expansion and/or radial-dependent acceleration) of the inner knots may 
be not that simple as assumed for a circular, 2D expanding ring.  

\begin{table}
\begin{center}
\caption{Kinematics of the inner knots of A30 and A78}
\label{tab:summary}
\begin{tabular}{lccrcc}
\hline
\hline
\multicolumn{1}{l}{Object} & 
\multicolumn{4}{c}{\underline{~~~~~~~~~~~~~~~Polar Outflow~~~~~~~~~~~~~~~}} & 
\multicolumn{1}{c}{Equatorial Ring} \\
\multicolumn{1}{c}{} & 
\multicolumn{1}{c}{$v_{\rm proj}$} & 
\multicolumn{1}{c}{$v_{\rm rad}$} & 
\multicolumn{1}{c}{$\vert v\vert$} & 
\multicolumn{1}{c}{$i$} & 
\multicolumn{1}{c}{$v_{\rm proj}\tablenotemark{a}$} \\
\multicolumn{1}{c}{} & 
\multicolumn{1}{c}{(km~s$^{-1}$)} & 
\multicolumn{1}{c}{(km~s$^{-1}$)} & 
\multicolumn{1}{c}{(km~s$^{-1}$)} & 
\multicolumn{1}{c}{(\degr)} & 
\multicolumn{1}{c}{(km~s$^{-1}$)} \\
\hline
A30 & 62 & $-$28\tablenotemark{b}  &  68~~~ & 65.7 & 56 \\
A78 & 78 & $+$98\tablenotemark{c} & 125~~~ & 38.5 & 30 \\
\hline
\end{tabular}
\tablenotetext{a}{
Averaged values.
}
\tablenotetext{b}{
Adopted from \citet{Meaburn1996}.   
}
\tablenotetext{c}{
Adopted from \citet{Meaburn1998}. 
}
\end{center}
\end{table}

\section{Discussion} \label{section4}

If the knots and filaments of A30 or A78 were ejected in a single event and 
have moved ballistically since then, without further interactions, then all 
the knots should have similar $\delta_\mathrm{exp}$ and $\tau_\mathrm{exp}$.  
The observational evidence for a non-linear correlation of the expansion 
with radial distance suggests that the actual dynamics is far from simple.

There are two plausible dynamical origins for the H-poor knots.  
The first one is that 
the knots were ejected from the star during the born-again event or along its 
return to the AGB, in which case they could be expected to have high initial 
velocities.  The second one is that they formed as a result of instabilities 
during the early stages of the interaction of the new, accelerating fast wind 
with the recently expelled dense material.  Knots formed in this way could be 
expected to have low initial velocities, since they are a result of 
condensations forming in a slowly moving dense shell.  Material moving at a 
few hundreds km~s$^{-1}$ was detected in Sakurai's object only a few years 
after its outburst \citep{Kerber_etal2002}.  It cannot be assessed whether 
it corresponds to fast moving bullets or to a disrupted shell 
\citep{Hinkle2014}, but it clearly shows that some material can move at high 
speeds early in the evolution of a born-again PN.

However the knots are formed, their later dynamics is determined by 
how they interact with the stellar wind, the ionizing photon flux, and
the pre-existing nebular material in the original PN.  
We consider each of these mechanisms in turn, although the actual 
dynamics will be a combination of all the effects, with each process 
being likely prevalent at varying distances from the star.

We begin by considering the interaction with the stellar wind, which 
was suggested to be at the origin of the acceleration of these knots 
by \citet{bor95}.  
The fast wind from the CSPN is highly supersonic and rams into the 
born-again ejecta and into the slowly expanding (30--40 km~s$^{-1}$) 
old nebula.  
The knots can either be embedded in freely flowing fast wind or in the 
hot, shocked wind bubble. 
The interaction of a knot with the highly supersonic, free-flowing 
wind will lead to the formation of a bow-shock in the wind at the 
star-facing side of the knot, whereas another shock is transmitted 
into the dense knot material, compressing and heating it.  
Material is stripped from the knot in turbulent mixing layers, 
mass loading the stellar wind and modifying the dynamics of the 
downstream flow \citep{hdps1986}. 
If the knot is inside the hot, shocked wind bubble, the interaction occurs 
with a subsonic flow, but similarly turbulent mixing layers erode the knot 
material \citep{al1997} and mass-loading modifies the flow dynamics 
\citep{pittard05}.

During these interactions, a fraction of the wind momentum is transferred 
to the eroded material and another fraction accelerates the knot as a whole.  
In the free-flowing wind, the momentum flux transferred from the wind 
to the knot is 
\begin{equation}
\dot{\it P} = \epsilon \dot{M} v_\infty \frac{r_\mathrm{c}^2}{R^2}, 
\end{equation}
where $\dot{M}$ and $v_\infty$ are the mass-loss rate and terminal wind 
velocity, $R$ is the radial distance to the star, $r_\mathrm{c}$ is the 
knot radius, and $\epsilon \leq 1$ is the momentum transfer efficiency.  
For a knot embedded in the subsonic flow, the momentum flux transferred 
to the wind is 
\begin{equation}
\dot{\it P} = \epsilon \dot{M} v_\infty \frac{r_\mathrm{c}^2 
R_\mathrm{s}^2}{R^4} \frac{(\gamma-1)^2}{(\gamma+1)^2}, 
\end{equation}
where $R_\mathrm{s}$ is the radius of the stellar wind shock and $\gamma$ 
is the ratio of specific heats.  
Under the assumption that the knot radius is roughly constant, the momentum 
transfer drops off with radial distance from the CSPN both in the cases 
of the free-flowing wind and the shocked, subsonic flow.

The knots are also exposed to ionizing radiation, causing the characteristic 
elongated, straight shadow tails of knots revealed by the images of A30 and 
A78.  
This is a known effect which is particularly well 
illustrated by the numerous ``cometary'' knots in NGC\,7293, the Helix 
Nebula \citep[e.g.,][]{odell96,mb2010}, although we note that the cometary 
knots in A30 and A78 are much closer to the ionizing source than those in 
the Helix.
If the knots are dense enough to contain neutral material \citep{Evans2006}, 
then an ionization front will propagate into the knot, contributing to its 
dynamics as also proposed for the expansion of the knots in the Helix 
Nebula \citep{Meaburn1992,odell96}. 
The initial implosion stage and subsequent acceleration of the globule due 
to the rocket effect have been discussed in detail in a series of papers 
\citep[see e.g.,][and references therein]{ber1989,bm1990,hen2009}. 
Basically, the effect on a particular knot depends on its initial 
column density and the ionization parameter at the location of the 
knot.  
As for the innermost knots of A30 and A78, only modest velocities, $\leq$5 
km~s$^{-1}$, can be gained due to the rocket effect of the photoevaporating 
gas.  
Moreover, the ionization front advances in early stages into the neutral 
material of the knot at the sound speed, $c_\mathrm{n} = kT/\mu m_\mathrm{H}$, 
which corresponds to $\approx$3 km~s$^{-1}$ for H-poor neutral material at 
3000~K.  

The interaction of ionizing photons with the knots has another consequence: 
the neutral material is photoevaporated and flows away from the knot 
ionization front and back towards the ionizing source, i.e., the CSPN.  
This flow of photoevaporated material can reach velocities of $\sim30$ 
km~s$^{-1}$ and then shocks against the free-flowing wind or the subsonic, 
hot, shocked wind, producing a bow shock (as seen, for instance, in 
protoplanetary disks in Orion; \citealt{garcia2001,garcia2002}).  
The photoevaporated material will eventually mix with the stellar wind 
in a similar way to the hydrodynamic mass-loading processes described 
above.

The stellar wind (either free-flowing or shocked), the knots, and the 
material eroded from these knots and carried outwards by the stellar 
wind will ultimately interact with pre-existing material in the original 
PN.  
As the H-poor material travels away from the CSPN, it sweeps up material 
from the ambient nebula and loses speed.  
The density of material inside the old nebular shells of A30 and A78 
is expected to be low, as implied by their low surface brightness and 
limb-brightened morphology.  
Therefore, at small distances from the CSPN, the dynamical effects 
due to the interaction with the old PN are minimal.
As the H-poor material moves outwards, however, it will pick up more and 
more nebular material, and the initial momentum averaged over the total 
mass is reduced due to mass loading. 
We should expect this effect to be noticeable at large distances from 
the CSPN, with maximum impact on light parcels of gas moving at high 
speeds such as those detected in the high-dispersion spectra of A30 and 
A78 by \citet{Meaburn1996} and \citet{Meaburn1998}.

To explore the effects of the stellar wind, ionizing photons, and mass 
loading on the dynamics of the H-poor ejecta, we have first obtained 
analytical solutions of the motion of different clumps of material 
and then performed basic radiation-hydrodynamic simulations.  
These are described in Sections~\ref{section4}.1 and \ref{section4}.2, 
respectively.

\subsection{Analytical Models}

For our simple analytical models of the physical processes involved 
in the dynamics of the H-poor knots of A30 and A78, we have assumed that 
these knots can gain a fraction of the momentum available from the stellar 
wind, but they can also be accelerated by the rocket effect of the 
photoevaporating gas ($\leq$5 km~s$^{-1}$) or experience an apparent shift 
outwards in their radial position induced by the progression of the 
ionization front ($c_\mathrm{n} \leq 3$ km~s$^{-1}$) that mimics a radial 
expansion velocity.  
This positive (real or apparent) acceleration depends on the cross 
section of a knot and its mass, declining very steeply with its 
radial distance from the source of stellar wind and ionizing flux, 
i.e., the CSPN.  
Meanwhile, as a knot or a parcel of gas travels away from the CSPN, it 
can experience the ram pressure of the ambient nebula and, by sweeping 
up its material, the H-poor gas loses speed.  
Similarly, the deceleration suffered by a knot depends on its cross section 
and its mass, but in this case the effects increase with its distance from 
the CSPN.

The dynamical evolution of the H-poor knots and parcels of gas carried by 
the stellar wind has been modeled for a born-again event that occurred in 
1200 AD, i.e., 814~yr ago.
It was further assumed that the stellar UV flux and fast stellar wind 
turned on one hundred years after the born-again event, reaching stellar 
wind parameters as those of A30 and A78, $\dot{M}$$\sim$10$^{-8}$ 
$M_\odot$~yr$^{-1}$ and $v_\infty\lesssim$4000 km~s$^{-1}$ 
\citep{Guerrero2012,Toala2014}. 
A range of sizes (1.3$\times$10$^{16}$~cm $<r_\mathrm{c}<$
2.7$\times$10$^{17}$~cm) and initial masses ($m_0>$ 1$\times$10$^{28}$~g), 
as derived from the size and averaged density 
($\rho\sim$1.2$\times$10$^{-21}$~g~cm$^{-3}$) adopted for the knots, and 
initial velocities (10~km~s$^{-1}<v_0<$~400~km~s$^{-1}$) were assumed 
for the knots or parcels of gas carried away by the stellar wind. 
Similarly, a range of the momentum transfer efficiency, $10^{-5}<\epsilon<1$, 
was assumed.  As for the outer nebular shell, a constant density of 
100~cm$^{-3}$ was adopted.

\begin{figure}
\begin{center}
\includegraphics[width=0.95\columnwidth,angle=0]{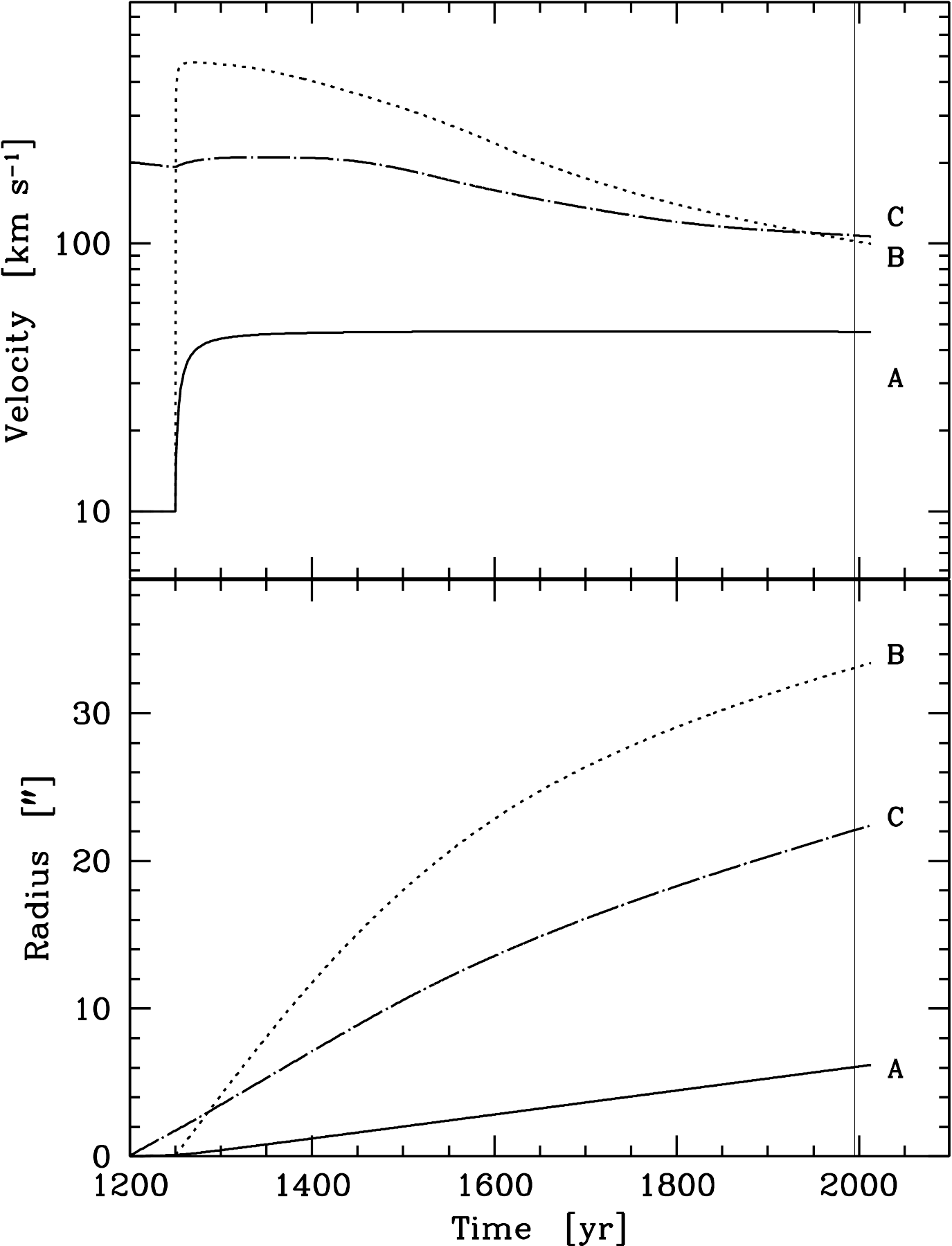}
\end{center}
\caption{
Time evolution of the expansion velocities (\emph{top}) and radial 
distances (\emph{bottom}) of different H-poor knots predicted by our 
analytical models for a born-again event which occurred 814~yr ago.  
Knot A (solid line) is heavy and initially slow (10~km\,s$^{-1}$), knot B 
(dotted line) has a similar low initial velocity, but it is light, and 
knot C (dot-dashed line) is also light, but it has a high initial 
velocity (200~km\,s$^{-1}$).  The vertical line indicates the time of the 
\emph{HST} WFPC2 observations, i.e., 1994/1995.  
}
\label{analytical}
\end{figure}

Our analytical model is illustrated in Figure~\ref{analytical}, which shows 
the evolution of different H-poor knots after the born-again event.  
The ejected material is pushed away by the stellar wind (and also 
photo-evaporated by the ionizing flux), and it is slowed down by 
the H-rich ambient material as it travels farther away from the CSPN.  
Heavy knots with 
low initial velocities (like knot A in Figure~\ref{analytical}) move 
relatively slowly, and its acceleration by the stellar wind is small.  The 
predicted fractional expansion ($\delta_{\rm exp}$) of such a knot is 2.5\%, 
which generally agrees with the observations (see Table~\ref{expansion}).  
The behavior of those knots can be compared to that of light parcels of 
H-poor gas eroded from heavy and slow-moving knots (like knot B in 
Figure~\ref{analytical}).  Such knots have low initial velocities, but 
they could be first significantly accelerated by the stellar wind and the 
ionizing flux, and then gradually decelerated due to their interaction 
with the nebular material.   The fractional expansion of knot B in 
Figure~\ref{analytical} is low, $\delta_{\rm exp}\approx$1.0\%, but its 
large radial distance from the CSPN indicates that it had very high 
velocities in the past.  
We have also considered the dynamical evolution of a light knot with high 
initial velocity (knot C in Figure~\ref{analytical}).  Since the high 
initial velocity of this knot ensures that it can travel farther from the 
CSPN than knots A and B by the time the stellar wind and ionizing flux of 
the CSPN are turned on, the stellar wind will have little effects on its 
dynamics.  
Soon after, the knot reached a distance where the dynamical effects of the 
ambient material dominate.  Consequently, its fractional expansion is low: 
for knot C, $\delta_{\rm exp}\approx$1.6\%.


The innermost knots in the equatorial disks have the highest values of 
fractional expansion $\delta_\mathrm{exp}$, implying that these knots 
have probably experienced a recent acceleration.  
Our models show that the stellar wind can sweep up small parcels of ejecta 
close to the CSPN at high speeds \citep{har1995}, but it cannot by itself 
produce the sharp decrease of $\delta_\mathrm{exp}$ with radial distance 
seen for the inner knots (Figure~\ref{fig3}).  
This discrepancy is lessened by the rocket effect of photoevaporating 
material and the progression of the ionization front: 
in the 20 years between observations, the rocket effect and progression 
of the ionization front into the knot will have moved outwards at the 
star-facing edge of the knots by $\sim5\times10^{14}$~cm, which corresponds 
to $\sim19$~mas at the distance of A30 and $\sim21$~mas at the distance of 
A78.  This corresponds to $\delta_\mathrm{exp}$$\sim$1\% at 2\arcsec\ and 
$\delta_\mathrm{exp}$$\sim$0.2\% at 10\arcsec\ from the CSPN, 
reflecting a quick drop in its contribution to $\delta_\mathrm{exp}$.

On the other hand, the lowest values of $\delta_\mathrm{exp}$ are associated
with the filamentary features immediately outside the equatorial ring of A30 
and the outermost filaments in sectors 1 and 2 of A78.  Knots B and C in 
Figure~\ref{analytical} semi-qualitatively describes those cases.  The low 
values of $\delta_\mathrm{exp}$ suggest a sudden deceleration, which 
is particularly intriguing for the outer filaments in A78 because their large 
radial distance implies that this material sustained a high expansion velocity 
in the past.  
Our models confirm that light (maybe also low-density) parcels of the H-poor 
material can be accelerated to high speeds by the stellar wind and travel 
quickly across the cavity up to the inner wall of the outermost nebular shell 
where they can experience a sudden deceleration.  
Such sharp decrease in velocity would result in the velocity spikes described 
by \citet{Meaburn1996} and \citet{Meaburn1998}, with their lowest expansion 
velocities being similar to the radial velocity of the H-rich shell.

\subsection{Numerical Simulations}

To assess in depth the effects of the physical processes described 
above, 2D radiation-hydrodynamical simulations of 
the formation of a born-again PN have also been carried out. 
However, the simulations and the results presented in this section 
are not meant to be an exact model of A30 or A78 but just to 
illustrate the complexity of the dynamics. 
The code described in \citet{TA2011,TA2014} was used and the interested 
reader is referred to those articles for a description of the numerical 
details. 
The 2D cylindrically symmetric calculations are performed on a fixed grid 
of 1000 radial by 2000 $z$-direction cells of uniform cell size and total 
grid spatial size of 0.5$\times$1.0~pc$^{2}$.  
The free-wind injection zone has a radius of 40 cells, corresponding to 
the innermost 0.02~pc.

\begin{figure}[!t]
\begin{center}
\includegraphics[width=1.0\columnwidth]{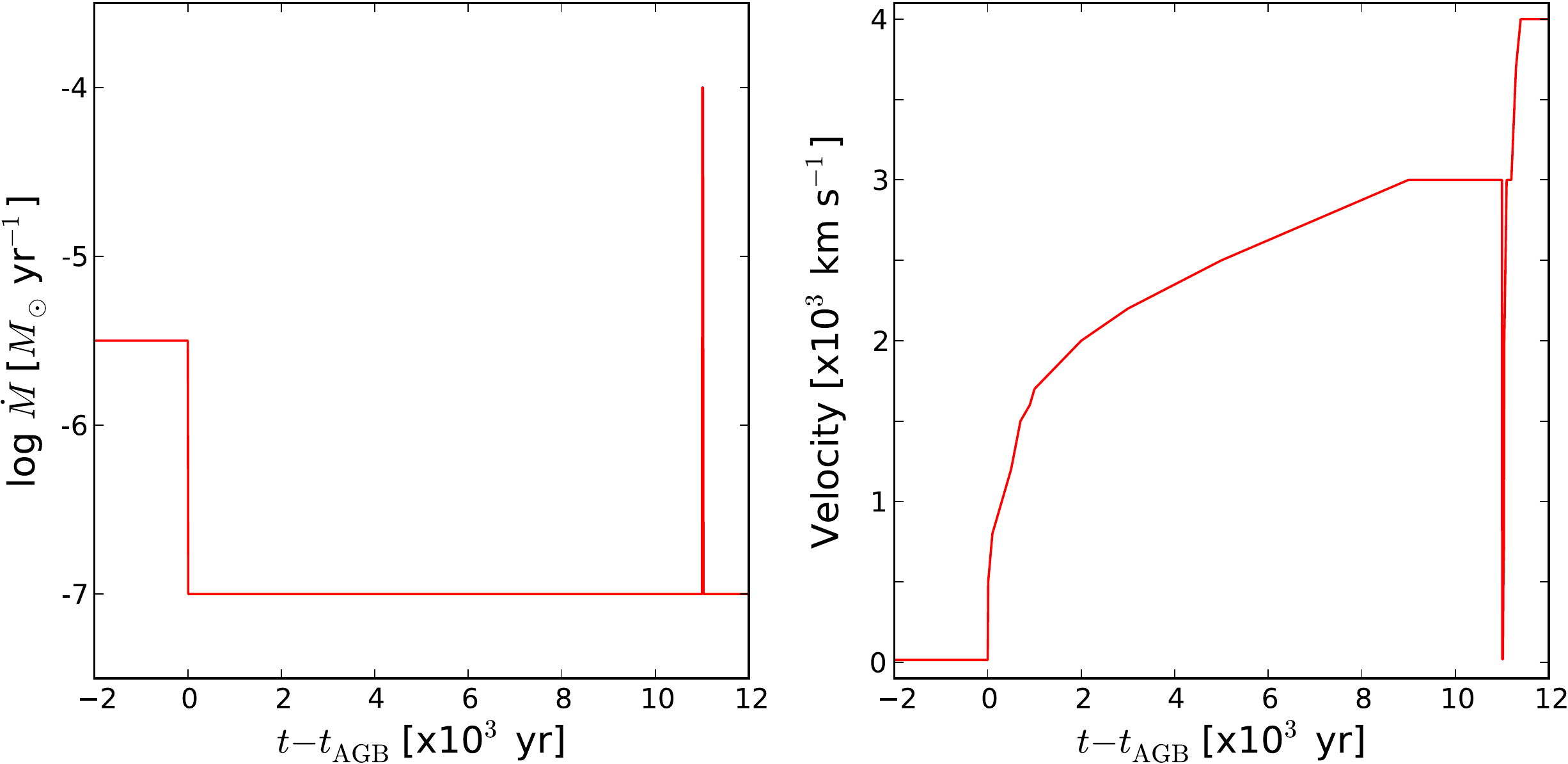}
\end{center}
\caption{Mass-loss rate and wind velocity profiles used for the simulations 
of the born-again PN.  The initial time is set at the end of the final 
phases of the AGB, whose duration is assumed to be 10$^{5}$~yr.  
}
\label{fig5}
\end{figure}

The numerical simulations have followed the evolution of the star through 
the final phases of the AGB and the post-AGB phases, then the born-again 
event, and finally along its loop through the AGB locus in the HR diagram 
and its current CSPN phase.  The initial conditions for the ISM were a 
number density $n_\mathrm{0}$=1~cm$^{-3}$ and a temperature of $T_{0}$ =
100~K.  Typical values for the duration of the AGB and the first CSPN 
phase, mass-loss rate, and terminal wind velocity were assumed as listed 
in Table~\ref{tab:tabla_jesus} and shown in Figure~\ref{fig5}.  
These same parameters are mostly unknown for the born-again episode, largely 
varying for different model assumptions \citep[e.g.,][]{MillerBertolami2006}.  
For simplicity, the born-again event will be described by a short phase 
($\sim$20~yr) of high mass-loss rate (10$^{-4}$ $M_\odot$~yr$^{-1}$) at 
low speed (20~km\,s$^{-1}$).  
After the born-again event, the star goes back to the post-AGB track for a 
second time and develops a fast stellar wind and a high ionizing photon 
flux.  The stellar wind velocity can be adapted to increase very rapidly 
after the born-again event according to observations towards Sakurai's 
object and A58 \citep[see e.g.,][]{Clayton2006,Hinkle2014}.  
We will assume that as soon as 30~yr after the born-again event, the 
wind velocity reaches values as high as 3000--4000 km~s$^{-1}$ as seen 
in A30 \citep[see][and references therein]{Guerrero2012} with values 
for the mass-loss rate similar to those immediately before the born-again 
event.  
The ionizing photon flux for both CSPN phases is assumed to be the same, 
$S_{\star}$ = 10$^{47}$~s$^{-1}$, while no ionizing photon flux is 
considered during the AGB and born-again phases.

\begin{table}
\begin{center}
\caption{Stellar parameters used in the simulation}
\label{tab:tabla_jesus}
\begin{tabular}{lccc}
\hline
\hline
\rule[-1mm]{0mm}{5.0mm}
Phase & Duration & $\log$($\dot{M}$) & $v_{\infty}$ \\
     &  (yr)    & ($M_{\odot}$~yr$^{-1}$) & (km~s$^{-1}$)\\
\hline
\rule[-1mm]{0mm}{5.0mm}
AGB   & 10$^{5}$ & $-$5.5           & 15 \\
CSPN$_{1}$    & 1.1$\times$10$^{4}$ & $-$7 & $\lesssim$3000 \\
Born-again  & 20    & $-$4           & 20 \\
CSPN$_{2}$    & \dots & $-$7           & 4000 \\
\hline
\end{tabular}
\end{center}
\end{table}

The first 1.11$\times$10$^{5}$~yr of evolution in our simulation corresponds 
to the formation of the old PN.  First, the AGB material interacts with the 
ISM, leaving a $\rho(r)$$\sim{r}^{-2}$ density profile 
\citep[e.g.,][]{Villaver2002,Perinotto2004}.  Here we assume a constant 
mass-loss rate and a constant wind velocity.  
Then the star evolves through the post-AGB phase (CSPN$_{1}$ in 
Table~\ref{tab:tabla_jesus}) and its stellar wind sweeps up the dense AGB 
material whereas it is ionized by the high photon flux.  We let the PN 
evolve for 11\,000~yr, which is comparable with the dynamical ages of A30 
and A78 \citep{Meaburn1998,Guerrero2012}.

At this point, the circumstellar medium around the CSPN has the following 
configuration:  a free streaming wind region, a bubble filled of tenuous 
hot gas ($T\geqslant$10$^{7}$~K, $n\lesssim$10$^{-2}$~cm$^{-3}$), and the 
outer ionized PN.  The born-again event is set up at this moment.  It results 
in the ejection of dense and slow H-poor material inside the hot bubble that 
will be swept up as soon as the star enters the post-AGB phase for a second 
time (CSPN$_{2}$ in Table~\ref{tab:tabla_jesus}). 

\begin{figure*}[ht]
  \begin{adjustbox}{addcode={\begin{minipage}{\width}}{\caption{
\label{fig6}
  Total ionized number density (\emph{left}), temperature
  (\emph{middle}), and velocity (\emph{right}) for the evolution of
  a PN after the born-again event. The two rows correspond to different
  times marked on top of each panel.
      }\end{minipage}},rotate=90,center}
      \includegraphics[scale=.9]{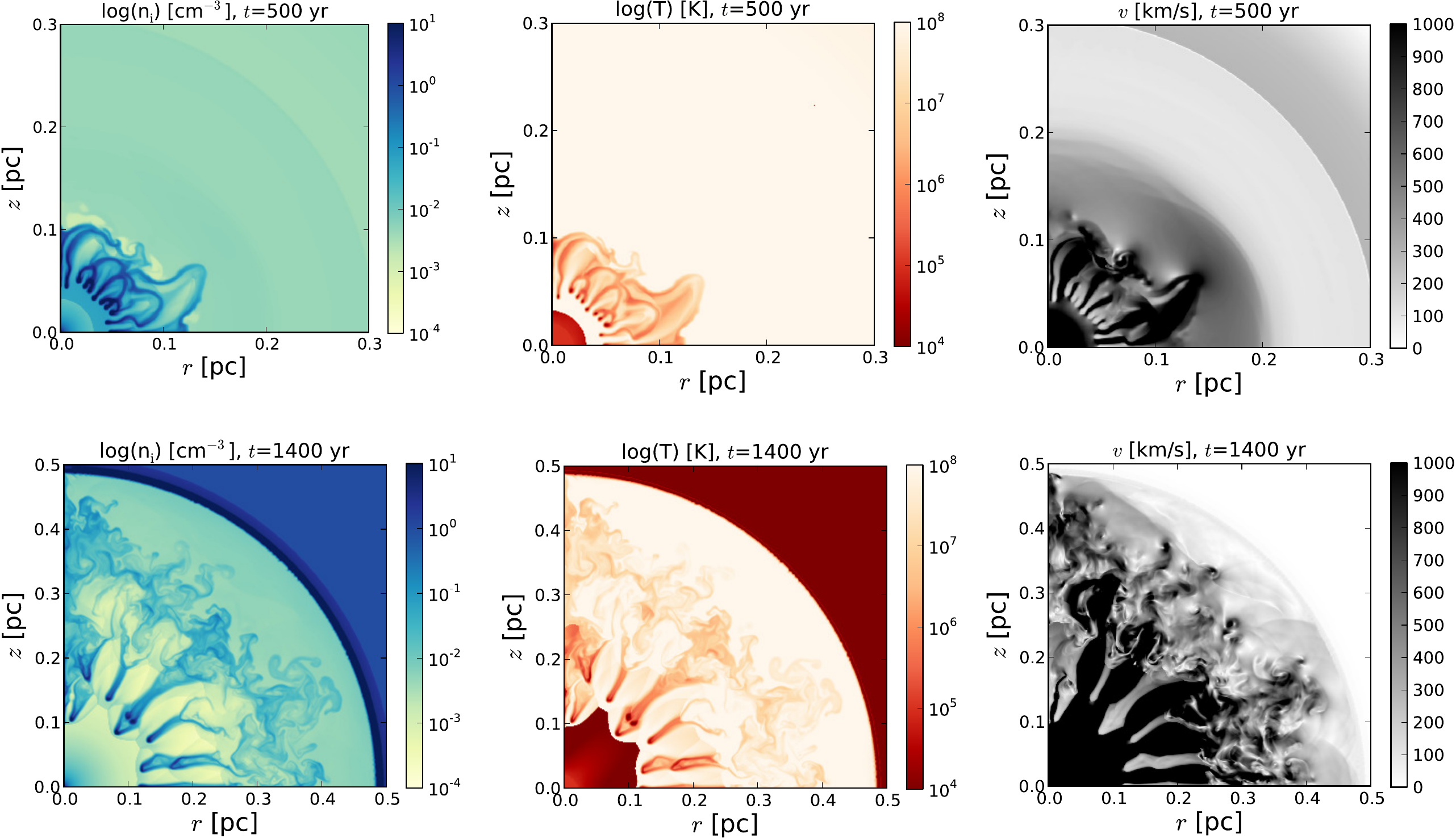}
  \end{adjustbox}
\end{figure*}

The time evolution of the ionized number density ($n_\mathrm{i}$), 
temperature ($T$), and radial velocity with respect to the central star 
($v$; $v^{2}=v_{r}^{2}+v_{z}^{2}$, where $v_{r}$ and $v_{z}$ are the radial 
and vertical velocities on each cell, respectively) after the born-again 
event are presented in Figure~\ref{fig6}.  
At early times, the star ionizes the dense material left by the born-again 
ejection whilst the fast wind sweeps it up, creating hydrodynamical 
instabilities in this interaction.  
After 500~yr of the onset of the CSPN phase (Figure~\ref{fig6}-\emph{top}), 
the fast stellar wind has disrupted the shell left by the born-again ejection 
and photoionized clumps are formed.  
These clumps have cometary shape, with dense, cold (10$^{4}$~K) heads 
pointing toward the central star and long tails that extend up to large 
radial distances because their velocities have been increased by the 
interaction of material in the clumps with the shocked fast wind.  
Multiple interacting shocks develop, heating the ablated clump gas at 
temperatures that can cause the X-ray emission detected toward A30 and 
A78 \citep{chu95,Guerrero2012,Toala2014}.

After 1400~yr (Figure~\ref{fig6}-\emph{bottom}), the 
densest clumps are still close to the star while the material that has 
been eroded fills regions of the old PN.  
Ejecta material carried away by the shocked stellar wind is interacting 
with the outer, old PN, suffering a strong deceleration.  
The final configuration of our simulations is thus very similar to that 
discussed in Section~\ref{section3}, showing material at different 
distances from the CSPN as a result of a complex dynamical evolution.

Our simulations show how a single and short born-again event is capable of 
producing the intricate morphology seen in A30 and A78, but it is important 
to recall that the evolution of the stellar and wind parameters of the CSPN 
before, during, and after the born-again event has been largely simplified.  
The situation is especially critical for the born-again event itself, 
for which a reliable temporal evolution of the stellar wind parameters 
is missing.  
Furthermore, the initial born-again ejecta is assumed here to be 
isotropic, whereas the distribution of the central clumps in all 
born-again PNe suggests otherwise.  
Finally, the cooling rate of the material has not been tailored to 
the peculiar chemical abundances of the H-poor ejecta.  
More detailed simulations on the formation, evolution, and X-ray emission 
from these objects will be presented in a forthcoming paper (Toal\'{a} \& 
Arthur, in preparation).

\subsection{Expansion Ages}

Figure~\ref{fig3} clearly shows that there is not a unique apparent 
expansion age for A30 or A78.
The outermost H-poor filaments must have been slowed down in their
interaction with the outer H-rich shell, thus the apparent expansion age 
derived from these features is overestimated.

On the other hand, the equatorial rings are affected by erosion and
ionization-front propagation.
The average expansion age of 735$^{+185}_{-125}$~yr derived for the
equatorial knots of A30 and 705$^{+135}_{-100}$~yr for those of A78
must thus be considered as lower limits.  

On the contrary, the polar outflows seem to have been slowed down.
The expansion age of 800$^{+150}_{-110}$~yr and $\sim$1140~yr implied from 
the polar knots of A30 and A78, respectively, must be thus regarded as 
upper limits.

To conclude, the dynamical effects suffered by the different structural 
components of A30 and A78 hinder the determination of the age of the 
born-again event.  
The expansion age of equatorial rings and polar outflows bracket the time 
since the born-again event as 610--950~yr for A30 and 605--1140~yr for A78.

\section{Concluding Remarks} \label{section5}

Multi-epoch \emph{HST} narrowband [O~{\sc iii}] images of the born-again 
PNe A30 and A78 obtained almost 20 yr apart reveal the angular expansion 
of their H-poor knots and filaments.  
The angular expansion of individual features generally increases 
with radial distance from the CSPN, but there is not a tight linear 
correlation and the expansion is evidently non-homologous.  
This argues against the ballistic expansion expected after 
a single ejection episode.

Hydrodynamical effects are important, as emphasized in illustrative 
simulations, 
and the current morphologies and structures of the H-poor material in A30 
and A78 are caused by interactions with the stellar wind (free-flowing or 
shocked), the ionizing photon flux, and the pre-existing (H-rich) nebular 
material.
The high fractional expansion of the H-poor knots at the equatorial 
rings seems to be best interpreted as the effect of erosion by the 
ionizing flux and fast stellar wind, rather than a dynamical acceleration.  
On the other hand, the azimuthal variation of the fractional expansion 
for the outermost features suggests that the stellar wind is channeled 
along preferred directions.

The expansion velocities of the equatorial rings and polar knots have 
been assessed.  
The inclination of the polar knots is not completely consistent 
with the aspect ratios of the equatorial rings assumed to be 
circular.  
The polar knots of A78, farther away from the CSPN than those in A30, 
expand faster, but their relatively low fractional expansion is indicative 
of recent deceleration.

Our panoramic investigation of the kinematics of the two born-again PNe 
complements previous morphological and spatio-kinematical studies. 
The current fast stellar wind ablates material from the H-poor knots, 
which is swept up downstream until it shocks the inner edges of the 
outer, H-rich nebula.  
The material carried by the wind suffers from a sharp deceleration 
and forms irregular shells (petal-like in A30, spindle-shaped in A78) 
as the stellar wind becomes anisotropic in its interaction with the 
inner knots.

\section*{Acknowledgements}
Support for the \emph{Hubble Space Telescope} Cycle\,20 General Observer 
Program 12935 was provided by NASA through grant HST-GO-12935.01-A from 
the Space Telescope Science Institute, which is operated by the Association 
of Universities for Research in Astronomy, Inc., under NASA contract 
NAS\,5-26555.
XF, MAG, RAML, and JAT are supported by the Spanish MICINN (Ministerio de 
Ciencia e Innovaci\'on) grant AYA~2011-29754-C03-02 co-funded with FEDER 
funds. 
JAT acknowledges support by the CSIC JAE-Pre~student grant 2011-00189.
RAML acknowledges support by Mexican CONACYT (Consejo Nacional de Ciencia y 
Tecnolog\'{i}a) grant NO.~207706. 
LMO acknowledges support from DLR grant 50\,OR\,1302.
SJA thanks DGAPA, UNAM for support through project PAPIIT~IN101713. 
We would like to thank Panayotis Boumis, referee of this paper, for the 
valuable comments and suggestions.

\end{document}